\documentclass{PoS}

\title{SU(3)-breaking corrections to the baryon-octet magnetic moments in chiral perturbation theory}

\ShortTitle{Baryon-octet magnetic moments in $\chi$PT}

\author{\speaker{J. Martin Camalich}\\
        Departamento de F\'{\i}sica Te\'orica and IFIC, Centro
Mixto Universidad de
Valencia-CSIC,\\ Institutos de Investigaci\'on de Paterna, 46071-Valencia, SPAIN\\
        E-mail: \email{camalich@ific.uv.es}}

\author{L. Alvarez-Ruso\\
Departamento de F\'isica, Universidad de Murcia, E-30071 Murcia, SPAIN
      \\
        E-mail: \email{luis.alvarez@ific.uv.es}}

\author{L.S. Geng\\
      Departamento de F\'{\i}sica Te\'orica and IFIC, Centro
Mixto Universidad de
Valencia-CSIC,\\ Institutos de Investigaci\'on de Paterna, 46071-Valencia, SPAIN\\
        E-mail: \email{lsgeng@ific.uv.es}}

\author{M.J. Vicente Vacas\\
      Departamento de F\'{\i}sica Te\'orica and IFIC, Centro
Mixto Universidad de
Valencia-CSIC,\\ Institutos de Investigaci\'on de Paterna, 46071-Valencia, SPAIN\\
        E-mail: \email{vicente@ific.uv.es}}

\abstract{We report a calculation of the baryon magnetic moments using covariant chiral perturbation theory within the extended-on-mass-shell renormalization scheme including intermediate octet and decuplet contributions. By fitting the two available low-energy constants, we improve the Coleman-Glashow description of the data when we include the leading SU(3) breaking effects coming from the lowest-order loops. We compare with previous attempts at the same order using heavy-baryon and covariant infrared chiral perturbation theory, and discuss the source of the differences.}

\FullConference{International Workshop on Effective Field Theories: from the pion to the upsilon \\
		February 2-6 2009\\
		Valencia, Spain}

\begin{document}

\section{Introduction}

In the limit that SU(3) is an exact flavor symmetry it is possible to relate the magnetic moments of the baryon-octet and the $\Lambda\Sigma^0$ transition to those of the proton and the neutron. These are the celebrated Coleman-Glashow formulas~\cite{Coleman:1961jn}. The improvement of this description requires the inclusion of a realistic SU(3)-breaking mechanism. In the last decades several calculations of the SU(3)-breaking corrections using $\chi$PT~\cite{Gasser:1984gg,Gasser:1987rb,Scherer:2002tk} have been performed. Most of the calculations have been done in the context of heavy baryon (HB) $\chi$PT~\cite{Jenkins:1990jv}, with~\cite{Jenkins:1992pi,Durand:1997ya} and without~\cite{Caldi:1974ta,Meissner:1997hn} the explicit inclusion of the decuplet resonances. In the HB approach it is necessary to work up to next-to-next-to-leading order (NNLO) to find a good agreement with data, although at the price of loosing the predictive power of the theory since at this level one has seven unknown low energy constants (LECs) to describe eight measured quantities. 

Calculations of the baryon-octet magnetic moments in covariant $\chi$PT have become possible only after the advent of the infrared (IR)~\cite{Becher:1999he} and the extended on-mass-shell (EOMS)~\cite{Fuchs:2003qc} renormalization prescriptions. In the IR approach~\cite{Kubis:2000aa} at next-to-leading order (NLO) the description is even worse than in the HB approach and it becomes necessary to reach NNLO. In contrast, we have found that within the EOMS scheme up to NLO the agreement with the data is not only better than in HB and IR but also than the Coleman-Glashow description~\cite{Geng:2008mf}. Moreover, it has been shown that this agreement with the experimental data is preserved even after one considers the contributions of the decuplet resonances~\cite{Geng:2009hh}. The inclusion of these contributions in SU(3)-flavor $\chi$PT is necessary since we deal with perturbations in $m_K$ and $m_\eta$ that are well over the typical scale for the onset of the low-lying decuplet resonances, $\delta=M_D-M_B\sim0.3$ GeV. Finally, it is worth mentioning that the covariant formalism successfully applied to describe the baryon-octet magnetic moments has been recently used to predict $f_1(0)$~\cite{Geng:2009ik}. 

\section{Chiral Lagrangians}

Our calculation requires the use of the standard lowest-order Chiral Lagrangians $\mathcal{L}^{(2)}_\phi$ and $\mathcal{L}^{(1)}_{\phi B}$, describing the pseudoscalar mesons and baryons coupled to an external electromagnetic source (see for instance~\cite{Scherer:2002tk}). Besides, for the meson-octet-decuplet interaction we use the ``consistent'' formalism that demands these couplings to fullfil a spin-3/2 gauge symmetry~\cite{Pascalutsa:2000kd,Pascalutsa:2006up}:
\begin{equation}
\mathcal{L}\,'^{\,(1)}_{\phi B D}=\frac{i\,\mathcal{C}}{M_D F_\phi}\;\varepsilon^{abc}\left(\partial_\alpha\bar{T}^{ade}_\mu\right)\gamma^{\alpha\mu\nu}
 B^e_c\,\partial_\nu\phi^d_b+{\rm h.c.},\label{Eq:CnsLag}
\end{equation}
where $T_\mu\equiv T^{ade}_\mu$ collects the field representation of the decuplet-resonances as detailed in ~\cite{Geng:2009hh}, $\mathcal{C}$ is the $MBD$ coupling, $F_\phi$ is the meson decay constant and $M_D$ is the decuplet-baryon mass. 

At second order there are two terms in the Chiral Lagrangian that contribute to the magnetic moments of the octet baryons
\begin{equation}
\mathcal{L}_{\gamma B}^{(2)}=\frac{b_6^D}{8 M_B}\langle\bar{B}\sigma^{\mu\nu}\lbrace F^+_{\mu\nu},B\rbrace\rangle+\frac{b_6^F}{8 M_B}\langle\bar{B}\sigma^{\mu\nu}[F^+_{\mu\nu},B]\rangle, \label{eq:BaryonLag2}
\end{equation}
where, in our case, $F^+_{\mu\nu}=2 |e| Q F_{\mu\nu}$, and $F_{\mu\nu}=\partial_\mu A_\nu-\partial_\nu A_\mu$ is the electromagnetic strength tensor. The LECs  $b_6^D$ and $b_6^F$ encode information about short-distance physics and should be determined from experiment within a given renormalization scheme. 

\section{Parameter values}

We take the values $D=0.80$ and $F=0.46$ for the meson-baryon couplings appearing in $\mathcal{L}^{(1)}_{\phi B}$, the value $\mathcal{C}\approx1.0$ for the $\phi B D$ coupling~\cite{Geng:2009hh} and we use an averaged meson decay constant $F_\phi\equiv1.17f_\pi$ with $f_\pi=92.4$ MeV. For the masses of the pseudoscalar mesons we take
$m_\pi\equiv m_{\pi^\pm}=0.13957$ GeV, $m_K\equiv m_{K^\pm}=0.49368$ GeV, 
$m_\eta=0.5475$ GeV while for the baryon masses we use the average among the members of the respective SU(3)-multiplets, $M_B=1.151$ GeV and $M_D=1.382$ GeV.

\section{Results}

\begin{figure}[t]
\includegraphics[width=\columnwidth]{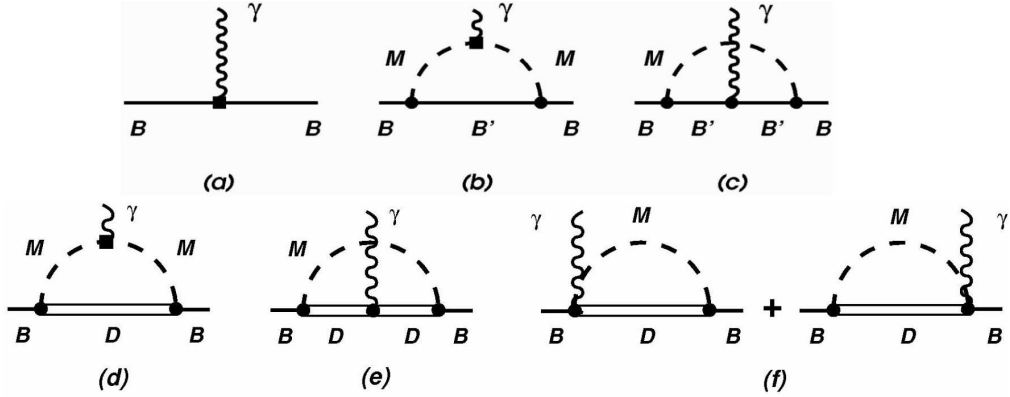}
\caption{Feynman diagrams contributing to the baryon-octet anomalous magnetic moments up to $\mathcal{O}(p^3)$. The solid lines correspond to octet-baryons, the double lines to decuplet-resonances, the dashed lines to mesons and the wiggly line denotes the external photon field. Black dots and boxes indicate $\mathcal{O}(p)$ and $\mathcal{O}(p^2)$ couplings respectively. \label{Fig:diagram}}
\end{figure}

The Feynman diagrams contributing to the baryon-octet anomalous magnetic moments up to $\mathcal{O}(p^3)$ are shown in Fig.~\ref{Fig:diagram}. The resulting expressions with detailed discussions on different regularization schemes of the loop-functions and on their heavy-baryon limits can be found in Refs.~\cite{Geng:2008mf,Geng:2009hh}. Furthermore,  Ref.~\cite{Geng:2009hh} analyzes the problems one may encounter in the relativistic field description of spin-3/2 particles.

\begin{figure}[t]
\hspace{2cm}\includegraphics[width=10cm]{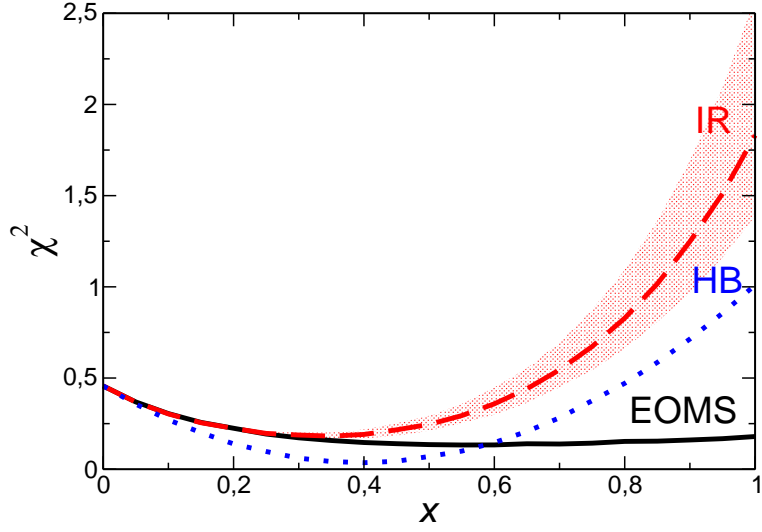}
\caption{(Color on-line) SU(3)-breaking evolution (see text for details) of the minimal $\bar{\chi}^2$ in the $\mathcal{O}(p^3)$ $\chi$PT renormalization of the approaches with only dynamical octet-baryons reported in this work. We also show the shaded areas produced by the uncertainty in $M_B$ when varying from 0.8 GeV to 1.1 GeV and chosing as central value $M_B=0.94$ GeV~\cite{Geng:2008mf}. This effect lies within the line thickness in the EOMS case, while the HB is insensitive to it.  \label{fig_graph}}
\end{figure}

\begin{table*}
\centering
\caption{Baryon octet magnetic moments in chiral perturbation theory up to $\mathcal{O}(p^3)$. We compare the SU(3)-symmetric description with the different $\mathcal{O}(p^3)$ $\chi$PT calculations discussed in the text. We display the HB and the EOMS-covariant results with (O+D) and without (O) the inclusion of dynamical decuplets and the IR-covariant with only dynamical baryons. We also include the experimental values from Ref.~\cite{Amsler:2008zzb}.  \label{Table:Results}}
\begin{tabular}{c|c|cc|c|cc|c|}
\cline{2-8}
& &\multicolumn{2}{|c|}{Heavy Baryon $\mathcal{O}(p^3)$}&Cov. IR $\mathcal{O}(p^3)$&\multicolumn{2}{|c|}{Cov. EOMS $\mathcal{O}(p^3)$}& \\
\cline{3-7}
&  \raisebox{1ex}[0pt]{$\mathcal{O}(p^2)$}& O & O+D& O &O &O+D & \raisebox{1ex}[0pt]{Expt.}\\ 
\hline\hline
\multicolumn{1}{|c|}{\textit{p}} & 2.56 &3.01& 3.47 &2.25& 2.60 & 2.61& 2.793(0)\\
\multicolumn{1}{|c|}{\textit{n}} & -1.60 &-2.62&-2.84& -2.67&-2.16 &-2.23 &-1.913(0)\\
\multicolumn{1}{|c|}{$\Lambda$}& -0.80 &-0.42&-0.17&-0.61&-0.64&-0.60&-0.613(4) \\
\multicolumn{1}{|c|}{$\Sigma^-$}& -0.97 &-1.35&-1.42&-1.15& -1.12 &-1.17&-1.160(25)\\
\multicolumn{1}{|c|}{$\Sigma^+$}& 2.56 &2.18&1.77&2.38& 2.41&2.37&2.458(10) \\
\multicolumn{1}{|c|}{$\Sigma^0$}& 0.80 &0.42&0.17&0.61& 0.64 & 0.60 & ... \\
\multicolumn{1}{|c|}{$\Xi^-$}& -1.60 &-0.70&-0.41&-1.12& -0.93 & -0.92 & -0.651(3) \\
\multicolumn{1}{|c|}{$\Xi^0$}&-0.97 &-0.52&-0.56&-1.28& -1.23 & -1.22 & -1.250(14) \\
\multicolumn{1}{|c|}{$\Lambda\Sigma^0$} & 1.38 &1.68&1.86& 1.88 &1.58&1.65& $\pm$1.61(8)\\
\hline
\multicolumn{1}{|c|}{$b_6^D$}& 2.40&4.71&5.88 &4.70& 3.92 & 4.30 & \\
\multicolumn{1}{|c|}{$b_6^F$}& 0.77 &2.48&2.49&0.43& 1.28 & 1.03 & ...\\
\multicolumn{1}{|c|}{$\chi^2$}&0.46&1.01&2.58&1.30& 0.18 & 0.22& \\
\hline
\end{tabular}
\end{table*}

In Table \ref{Table:Results} we show the numerical results for the baryon magnetic moments obtained by minimizing $\chi^2\equiv\sum(\mu_{th}-\mu_{expt})^2$ as a function of the LECs $b_6^D$ and $b_6^F$. We have not included the $\Lambda\Sigma^0$ transition moment in the fit and, therefore, it is a prediction.  We compare the SU(3)-symmetric description with the different $\mathcal{O}(p^3)$ $\chi$PT calculations discussed above. Namely, we display the HB and the EOMS-covariant results with (O+D) and without (O) the inclusion of dynamical decuplets and the IR-covariant with only dynamical octet-baryons. The experimental values are included from Ref.~\cite{Amsler:2008zzb}. 

For the HB approach, one sees how the corrections of the dynamical baryon -octet and -decuplet go in the same direction and are of equivalent size. Consequently, the description obtained with only the baryon-octet, that already overestimated the SU(3)-breaking corrections, gets much worsened when the decuplet is included. For the results obtained in HB, one is unavoidably led to wonder about the contributions of higher-mass resonances. In the covariant approach we have different results attending to the renormalization scheme or to the inclusion of the decuplet. 

Restricting ourselves first to the comparison between the EOMS and IR schemes with only dynamical baryon-octet contributions, we observe large numerical discrepancies for differences that are formally of higher-order~\cite{Geng:2008mf}. In order to better understand this, we study the evolution of the minimal $\chi^2$ as we {\it switch-on} the SU(3)-breaking effects, by introducing the parameter $x=m_M/m_{M,phys}$ (where $M=\pi, K, \eta$) and varying it between zero and one. As seen in Fig.\ref{fig_graph}, the three approaches coincide in the vicinity of the chiral limit. The EOMS and IR results stay very close up to $x\sim0.4$. As $x$ increases further HB and IR description of data get worse while, on the contrary, the EOMS result lies well below the SU(3) symmetric one. As it was explained in Ref.~\cite{Geng:2008mf}, we interpret the unrealistic IR behaviour as a manifestation of the alteration that this approach produces on the analytical properties of the loop-functions. Our results indicates that this known problem of the IR amplitudes may already be relevant for the scales around the masses of $K$ and $\eta$ mesons. 

The results in EOMS show an unprecedented NLO improvement over the tree-level description within dimensionally regularized $\chi$PT, where octet and decuplet contributions have been included. We can study the convergence properties of the chiral series factorizing the tree-level at $\mathcal{O}(p^2)$ from the whole result up to $\mathcal{O}(p^3)$. We also separate the loop fraction into the octet (second number) and the decuplet (third number) parts in the parenthesis
\begin{eqnarray*}
&&\mu_p=3.46(1-0.28+0.035)\hspace{0.15cm},\hspace{0.15cm}\mu_n=-2.86(1-0.16-0.06)\hspace{0.15cm},\hspace{0.15cm}\mu_\Lambda=-1.43(1-0.46-0.12),\\
&&\mu_{\Sigma^-}=-0.60(1+0.25+0.70)\hspace{0.15cm},\hspace{0.15cm}\mu_{\Sigma^+}=3.46(1-0.34+0.025)\hspace{0.15cm},\hspace{0.15cm}\mu_{\Sigma^0}=1.41(1-0.47-0.11),\\
&&\mu_{\Xi^-}=-0.60(1-0.07+0.61)\hspace{0.15cm},\hspace{0.15cm}\mu_{\Xi^0}=-2.86(1-0.48-0.09)\hspace{0.15cm},\hspace{0.15cm}\mu_{\Lambda\Sigma^0}=2.48(1-0.28-0.06).\label{Eq:convergence}
\end{eqnarray*}
Except for the $\Sigma^-$, the relative contributions of the octet and the decuplet and the overall $\mathcal{O}(p^3)$ corrections, are consistent with a maximal correction of about $m_\eta/\Lambda_{\chi SB}$, and the decuplet corrections are, in general, smaller than the octet ones. Moreover, in Ref.~\cite{Geng:2009hh} the large decuplet mass limit was investigated and it was found that the decuplet decouples for an average $M_D$ slightly above the physical average..  

Another interesting point concerns the different sum-rules first discussed by Caldi and Pagels in Ref.~\cite{Caldi:1974ta}. Among them, two survive up to the leading breaking corrections provided by any of the covariant $\chi$PT calculations considered. Namely, we found that our results verify
\begin{equation}
 \mu_{\Sigma^+}+\mu_{\Sigma^-}=-2\mu_\Lambda,\hspace{1cm}\mu_{\Lambda\Sigma^0}=\frac{1}{\sqrt{3}}\left(\mu_\Lambda-\mu_{\Xi^0}-\mu_n\right). \label{Eq:C-PSum-Rules}
\end{equation}
The first relation in combination with the assumed isospin symmetry is the cause of $\mu_\Lambda=-\mu_{\Sigma^0}$ in the results of Table \ref{Table:Results}. Experimentally, the two relations in Eq. (\ref{Eq:C-PSum-Rules}) are satisfied rather accurately, 1.298(27)=1.226(8) for the first relation and 1.61(8)=1.472(8) for the second. A combination of them produces the Okubo sum-rule~\cite{Okubo:1963zza}:
\begin{equation}
 \mu_{\Lambda\Sigma^0}=\frac{1}{2\sqrt{3}}\left(\mu_{\Sigma^0}+3\mu_{\Lambda}-2\mu_{\Xi^0}-2\mu_n\right). \label{Eq:Okubo}
\end{equation}
 The third sum-rule derived in~\cite{Caldi:1974ta}
\begin{equation}
\mu_{\Xi^-}+\mu_{\Xi^0}=2\mu_\Lambda-\mu_n-\mu_p, \label{Eq:3rdSR}
\end{equation}
 although fulfilled in the HB expansions of our results (see Ref.~\cite{Jenkins:1992pi}), is broken when the relativistic corrections to the loops are included. 

In summary, we have presented the leading SU(3)-breaking contributions to the baryon-octet magnetic moments in chiral perturbation theory considering different renormalization prescriptions and the explicit inclusion of the decuplet resonances. By using the so-called extended on-mass-shell renormalization prescription we successfully improve the tree-level description given long time ago by Coleman and Glashow. Our $\mathcal{O}(p^3)$ calculation intends to be complete in the sense that the low-lying decuplet resonances have been included. Finally, the comparison of our results with those obtained in heavy-baryon and infrared approaches highlights the important role that analyticity may have in baryon chiral perturbation theory. 
 
\section{Acknowledgments}

The authors thank J. Gegelia and V. Pascalutsa for useful discussions. This work was partially supported by the  MEC grant  FIS2006-03438 and the European Community-Research Infrastructure
Integrating Activity Study of Strongly Interacting Matter (Hadron-Physics2, Grant Agreement 227431) under the Seventh Framework Programme of EU. L.S.G. acknowledges support from the MICINN in the Program 
``Juan de la Cierva''. J.M.C. acknowledges the same institution for a FPU grant.

\end{document}